\documentclass[conference]{IEEEtran}
\IEEEoverridecommandlockouts
\usepackage{cite}
\usepackage{amsmath,amssymb,amsfonts}
\usepackage{algorithmic}
\usepackage{graphicx}
\usepackage{textcomp}
\usepackage{xcolor}
\usepackage{textcomp}

\def\BibTeX{{\rm B\kern-.05em{\sc i\kern-.025em b}\kern-.08em
    T\kern-.1667em\lower.7ex\hbox{E}\kern-.125emX}}
\begin{document}

\title{Exploring cultural challenges to implementing Educational Technology in the higher education sector in India }

\author{\IEEEauthorblockN{Parvathy Panicker}
\IEEEauthorblockA{\textit{ Department of Computing } \\
\textit{Goldsmiths University of London}\\
London, United Kingdom \\
ppani001@gold.ac.uk}
ORCID: 0000-0002-6934-6013

}

\maketitle

\begin{abstract}
When learning technologies are introduced in educational environments, it is assumed that the educational environment is culture neutral i.e, all educational environments have the same challenges, problems and cultural norms. However, it can be observed that cultural factors can influence the successful implementation and use of learning technologies. In this study the aims were to explore contextual challenges to implementing different educational technologies and to explore the effects of culture. The results of this survey suggest that Hofstede's cultural measures of uncertainty avoidance, power distance and Individualist/collectivist measures, and Duckworth's Grit measure of passion and perseverance have a strong impact on the culture of technology use in India. 
% The questions in this empirical study were designed to understand the specific cultural challenges, in order to amend the detailed classification methodology framework that could be implemented  for educational technologies in developing countries.
\end{abstract}
\section{Introduction}
Educational technology involves the integration of  technologies and media in instructional contexts and includes  integrating media into instructional processes to enhance the effectiveness of the teaching-learning process. Technology-enhanced learning platforms such as email, instant messaging, asynchronous discussion and synchronous conferencing are highly flexible and convenient, and their use can result in  powerful combinations of context, communication and purposeful activity which enhance the learning experience\cite{rawlins2014integrating}. One of the challenges of an intelligent educational technology platform is the ability to cater to a diverse cohort of students and teachers across different cultures \cite{PinkwartNiels2016A2oA}. Culture, referred to as the collective mental programming of the human mind\cite{HofstedeGeertH2010Cao:}, and education, which is the agent of culture, are inter-related \cite{BrameldTheodore1955CaE}: culture shapes pedagogy, educational policies and infrastructure and conversely, education is a major vehicle by which learners become enculturated\cite{AnsariDaniel2011Caen}.  The influence of culture on shaping educational policies and knowledge-sharing in the context of computer-based virtual classroom has already been highlighted \cite{ZhangXi2014Ceot}.  \\
Electronic learning is still in its infancy in developing countries, where the challenges in implementing  electronic learning platforms are different \cite{BhuasiriWannasiri2012CSFf}. Culture has been recognised as one of the conceptual challenges for implementing e-learning\cite{gronlund2010mobile},\cite{shraim2010learning}.  The implementation of educational technologies are influenced by unique cultural preferences in different societies. For example, the importance of including  images and symbols appropriate for the local culture has been  recognised\cite{andersson2009conceptual}. Language and interactive nature  contribute to culture and studies have highlighted the importance  of usefulness and perceived ease of use while using technology.\cite{ JungHee-Jung2015FaET}. For these reasons, transplanting technology-embedded platforms rendered in the setting of Western  culture would be inappropriate and  possibly irrelevant in a society where cultural pedagogy is different. \\
The aim of this study therefore is to explore different cultural factors that influence educational technology.\\
The study is divided into different  sections. After the present introduction, section II explains the background to this research. Section III sets out the experimental design and section IV describes the results of this survey. Section V presents the discussion, section VI sets out the conclusions and lastly section VII sets out the acknowledgement. 

\section{Background}

This section  briefly outlines some issues and concepts underpinning this research. The section begins by describing the (non)preference of technology in teaching, followed by a review of  fundamental challenges related to the use of electronic learning, and finally challenges  the effects of educational culture in adopting technology posed by a culture which values face-to-face presence of an instructor. \\ 

Traditional\lq chalk and talk\rq  teaching techniques is considered in many professions to be  superior for different reasons. 40.5\% of teachers prefer the traditional lecture delivery using chalkboard  \cite{boulos2007second}whereas some researchers argue that greater academic returns can be attained when instructors integrate technology such as mobile-learning, laptops and tablets \cite{castillo2017take}. However,  the rise of different technologies such as MOOC with open licensing, web based virtual learning environments, teleconferencing and more than 321 free educational technologies (elearningindustry.com) have made distance education readily accessible. The perceived ease of use and environmental context have been identified as determinants of e-learning adoption in universities whereas complex organisational compatibility can contribute to  e-learning neglect \cite{ansong2017determinants}. Institutionalising e-learning depends on student\textquotesingle s use of computers and  related technologies and hence e-learning tools should be made more attractive to be adopted by students \cite{boateng2016determinants}.\\
\cite{andersson2009conceptual} demonstrated that there are several challenges faced by institutions in developing countries when integrating  technology into learning platforms. The  need for quality assurance model for e-learning systems, absence of e-learning resources, lack of implementation process and the lack of instructional design have been identified as critical issues in localised environment\cite{FaridShahid2015Iapo}. Similarly, barriers such as electricity failure, supply\cite{oye2011challenges} and English proficiency were identified as most significant prevailing issues in Pakistani private Higher Education(HE) institutions\cite{qureshi2012challenges}. \cite{kamba2009problems} explored barriers faced by staff and students in Nigerian universities and observed that  internet related e-learning was used for finding information relating to research, web pages for advertisement of universities, transaction of students  rather than actual online learning. Importantly,  the study\cite{oye2011challenges} highlighted the problem of a narrow bandwidth impeding with the adoption of electronic learning. Likewise, barriers such as price, rural access and ICT literacy have been identified  in SriLanka\cite{gunawardana2005empirical}. Gunga and Ricketts\cite{gunga2007facing} have  identified that collaboration networks that include e-learning sponsors, policy makers, telecommunication network service providers and educators are required to solve problems around online education in Africa. Organisational challenges such as teacher training and social challenges including educational culture focused on  repetition\cite{gronlund2010mobile}are additional challenges. \\
 Even though ICT is considered as a facilitator of knowledge creation in leading economies(OECD, 1996),mindset\cite{tripathi2010learning} of stakeholders-students, administrators, teachers- have been identified as a barrier towards a flexible mode of teaching facilitated by e-learning. In fact, research demonstrates that technostress can be better acknowledged if the link between personality and culture along with technostress creators were understood\cite{krishnan2017personality}. Similarly, \cite{ZhangXi2014Ceot} have shown that cultural values(collectivism, concern for face) impact the nature of HE student\textquotesingle s knowledge sharing motivations. Importantly, factors such as culture and gender differences have been highlighted  in Chinese-US higher educational sector when using technology and suggest  that diversity within cultural groups among sub-cultures should be considered in understanding children\textquotesingle s IT use\cite{jackson2008culture}. Likewise,  differences in culture were recognised in a  Chinese-German study in which it was observed that the Chinese cohort preferred implicit communication styles from robots to accept recommendations  compared to the German cohort\cite{rau2009effects}. The factors influencing primary school teachers' acceptance of educational technology in China include facilitating conditions and attitude as the most significant\cite{wong2016behavioral}. These studies suggest that learning using technologies differ from country to country and depends on a number of factors that are unique to each situation- a proposition which this current study tries to elaborate.\\
Even though  cultural challenges  have been identified in  e-learning in the past, none of the studies have explored  the cultural effects in educational technologies. The detailed classification methodology framework\cite{hersh2017classification} identifies culture as one of the  factors in  personal characteristics in the ICT-based learning technologies for disabled people however, this  has been poorly explored. To address this gap, this study aims to understand the effects of culture in educational technologies and suggest a culturally inclusive framework for a cohort of faculty members in higher education institutions across India.\\

\section{Experimental design}
To evaluate the challenges faced by lecturers in the HE sector in India, the author carried out a cross-sectional study to evaluate the challenges of e-learning, cultural factors affecting educational technology and intrinsic motivation of lecturers to complete self-development and learning tasks. A questionnaire was designed covering demographic characteristics (gender, age, state , country of birth, area where brought up(town or village or city), highest education level(graduation, post graduation or PhD), current institution type(private-aided,  private-unaided or government), department,  years of teaching experience, years of teaching in the current institution and country of parent\textquotesingle s birth and their jobs) \cite{penelo2011factors}. Cultural measures  described by Hofstede (Power Distance(PD), Individualist/Collectivist(I/C), Masculine/Feminine(M/F) and Uncertainty Avoidance(UA) \cite{HofstedeGeertH2010Caos} and measures of  intrinsic motivation, in the grit scale (Passion(PA) and Perseverance(PE)) \cite{DuckworthAngelaL.2007GPaP} were included.  Lastly, open questions probing participants\textquotesingle views about technology were included. The questionnaire was administered to 3 participants and was amended based on their feedback. These questionnaires  were not included  in the  analysis.The final questionnaire consisted of 66 items distributed over 5 domains (key questions provided in appendix).  The responses to questions were measured using either using Likert  ( 5-point between 1- strongly agree /important/comfortable/likely/friendly and 5-strongly disagree /unimportant/uncomfortable/unlikely/unfriendly) or dichotomous  scaling (agree/disagree). The responses to questions for grit were measured using Likert scaling (5-points between 1-Not at all like me and 5-very much like me).
The survey was administered through  telephone,  paper or electronically during the Spring and Autumn semesters of 2018 between  March and September. Thrity-three institutions were randomly selected from Google's higher educational institutions in India and teaching staff were contacted through email by using addresses available on the respective universities' website.  Participants who were using educational technology were invited to participate after providing informed consent and they returned the completed questionnaires through email. Those contacted through telephone were taken through the structured questionnaire and responses were entered. Figure 1 depicts the inclusion of participants. Data were treated with strict confidentiality and anonymity.\\ 

\subsection{Statistical analysis}
Most statistics and graphical outputs were developed using R studio(Version 1.1.456) and one graph was developed using Microsoft Excel(Version 16.17).

\section{Results}
\subsection{Demographic characteristics}
\begin{figure}
\includegraphics[width=70mm, scale=0.5]{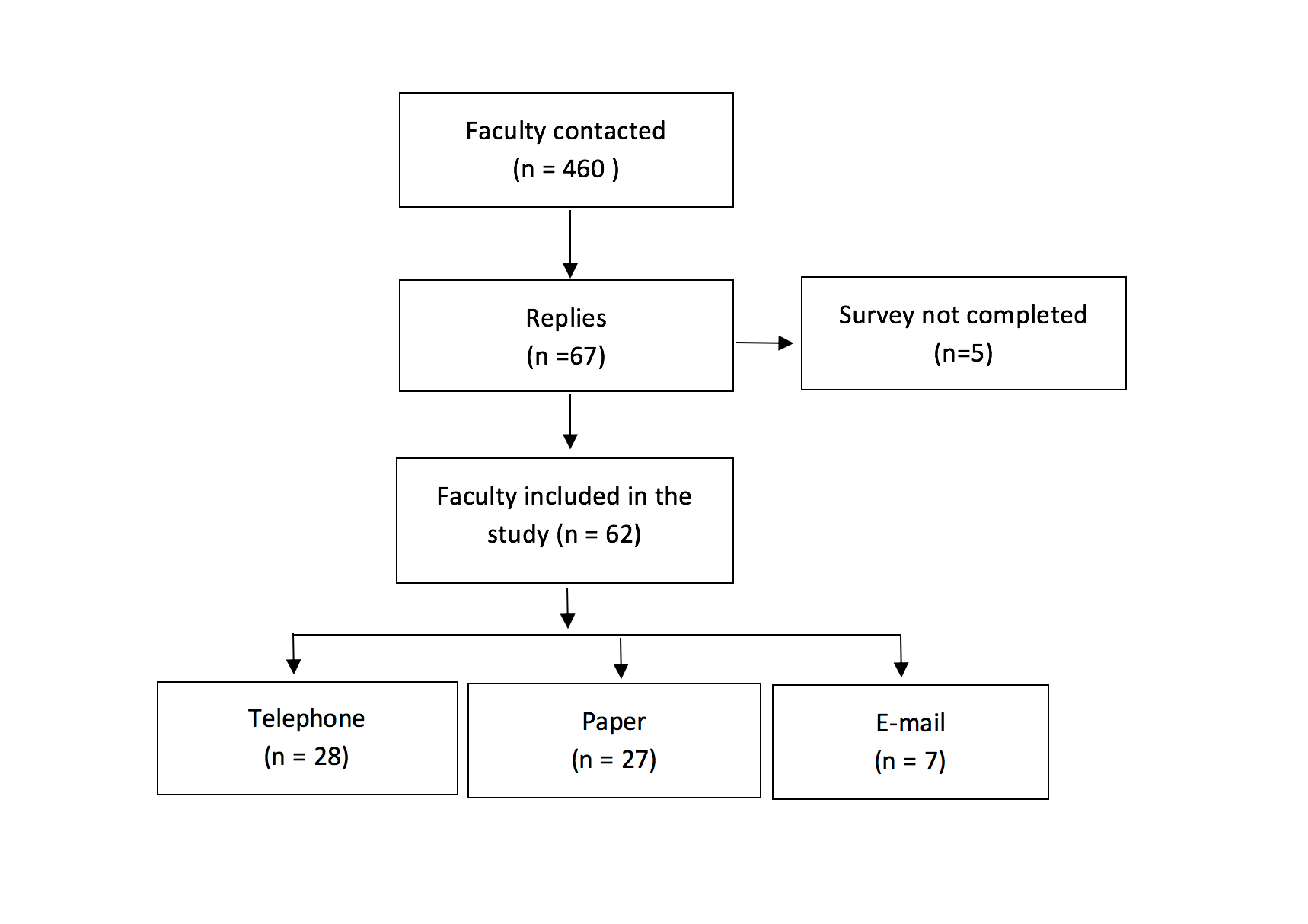}
  \caption{Flowchart of inclusion of participants}
\end{figure} 

Figure I illustrates the inclusion of participants in the study. The mean age of 62 participants was 44.22 and SD 9.56. Figure 3 illustrates the  age distribution of HE staff. 60\%(n=37) were females. Figure 2 illustrates teaching experience(mean duration 11years)

\begin{figure}
\includegraphics[width=70mm, scale=0.5]{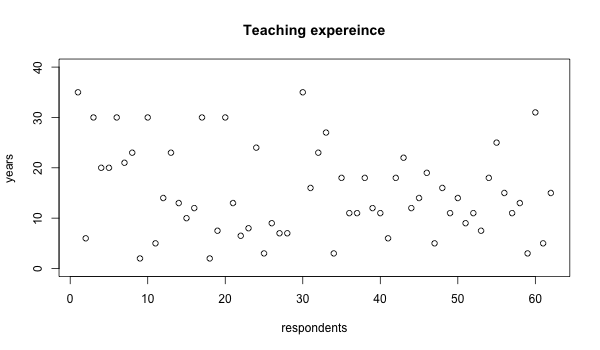}
  \caption{Teaching years Distribution}
\end{figure} 
\begin{figure}
\includegraphics[width=70mm, scale=0.5]{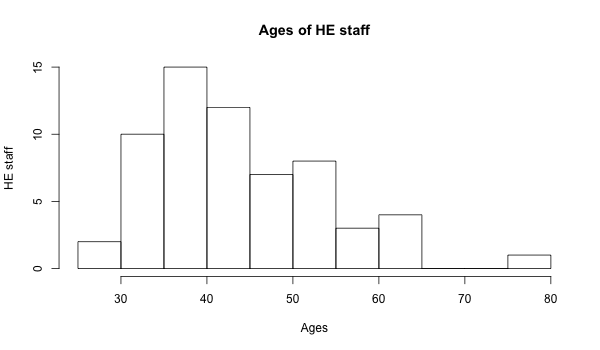}
  \caption{Age Distribution}
\end{figure}

Figure 4 depicts  the distribution of participants across HE institutions-government institutions 29\%(n=18), private-aided 6\%(n=4) and private-unaided 63\%(n=39). One participant worked at both government and private-unaided institutions. 
50\%(n=31) of respondents were holding a PhD and the remaining had only Masters degree 50\%(n=31)\\

\begin{figure}
\includegraphics[width=70mm, scale=0.5]{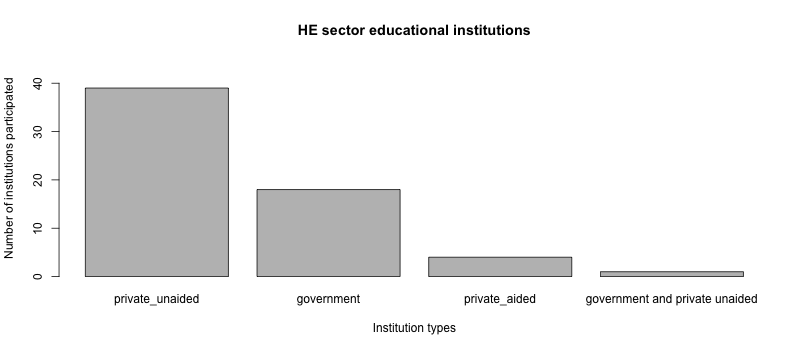}
  \caption{HE institutions}
\end{figure}

Table I shows the distribution of participants across specialities. 
 \begin{table}
 \begin{tabular}{ll}	\hline
\emph{Specialities} & \emph{Responses} \\ \hline 
Master of Comp Appl&16\\
Industr Eng and Magmt&12\\
Neurology&4\\
Computer Sc Eng&3\\
Pathology&2\\
Marketing&2\\
English&2\\
Zoology&2\\
Electronics and Comm Eng&1\\
Aerospace&1\\
Agriculture&1\\
Operations&1\\
Informations systems&1\\
Neurosurgery&1\\
Master of Busness App &1\\
Management&1\\
Telecom Eng&1\\
Child health&1\\
Physics&1\\
Electr and electro eng&1\\
Distance education&1\\
Chemistry&1\\
Optoelectronics&1\\
Dentistry&1\\
Oral Pathology&1\\
Pbl health and cmty medicine&1\\
Nursing&1\\
\end{tabular}\\
 \caption{\label{tab:table-name}Departments}
\end{table}

\subsection{Technological use}
The mode of teaching was either contact class (55\%(n=34)),  both contact classes  and web based teaching (34\%(n=21))  or web based teaching only (3\%(n=2)) (figure 5).  The commonest challenges to using technology for teaching were updating e-learning systems 42\%(n=26) , limited time in the teaching curriculum practice 39\%(n=24), limited bandwidth 35\%(n=22), obtaining appropriate hardware resources 32\%(n=20) and procuring relevant software including open source for teaching 31\%(n=19). Other challenges included a steep learning curve to use electronic learning 31\%(n=19), internet connectivity  29\%(n=18),  intermittent power supply 26\%(n=16), expenses involved 24\%(n=15),  limited staff expertise availability 24\%(n=15), confidence levels of students in using technology (21\%(n=13)), English competency of students  (17\% (n=11)) and the institution\'s policy around using technology (14\%(n=9)).  Figure 7 illustrates the different challenges reported. Other challenges that were reported in the open questions included :\\

\begin{enumerate}
\item Infrastructure facility to match student population in class, 70 students against 14-15 students in European universities
\item To hold attention of students
\item Websites can become slow
\item Audio can become blurry
\item Nature of methodology and techniques
\item No resources available to teacher
\item Time constraint when designing for the first time
\end{enumerate}

\begin{figure}
\includegraphics[width=70mm,scale=0.5]{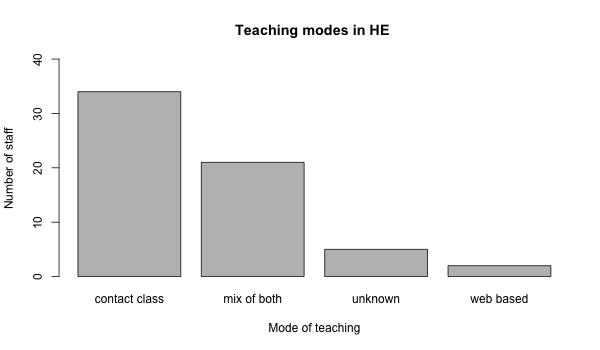}
  \caption{Teaching modes in teaching}
\end{figure}

 98\%(n=61) of the respondents stated that they used electronic resources when preparing for  lessons. and only 1\%(n=1)reported not using technology.\\ \\

Figure 6 depicts how technology and electronic resources were being used by faculty members for  lesson preparation. 90\% (n=56) of respondents used online resources such as PDFs, viewed online videos and read Wikipedia. 79\%(n=35) respondents used e-books when preparing, 56\%(n=35)used learning platforms such as Moodle or institutional in-house learning platforms,47\%(n=29)indicated they used CD ROMS while the remaining 40\%(n=25) indicated they had worked online to achieve an online qualification.

\begin{figure}
\includegraphics[width=70mm,scale=0.5]{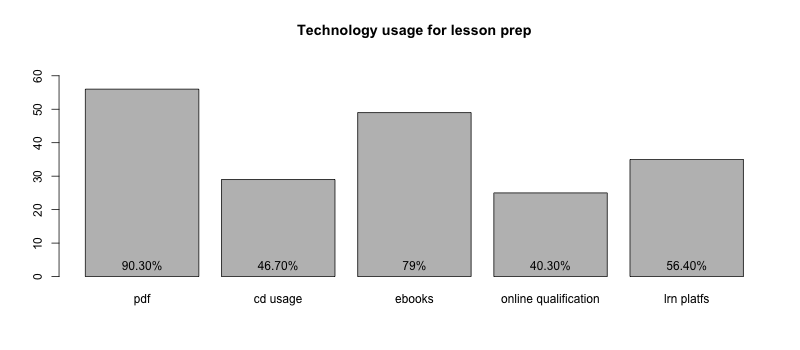}
  \caption{Technology use}
\end{figure}

45\%(n=28)used the internet in different forms whilst teaching.  43\%(n=27)said they did not make use of this facility. 11\%(n=7) did not answer this question in this survey.\\ \\

\begin{figure}
\includegraphics[width=100mm, scale=0.5]{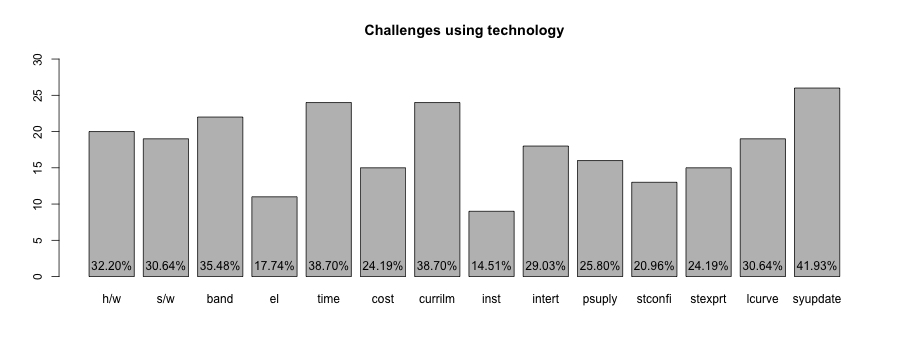}
  \caption{Technology challenges }
\end{figure}
Some of the \textbf{other challenges} reported as free text were :\\ 

\begin{itemize}
\item Infrastructure facility to match student population in class, 70 students against 14-15 students in European universities
\item To hold attention of students
\item Websites can become slow
\item Audio can become blurry
\item Nature of methodology and techniques
\item No resources available to teacher
\item Time constraint when designing for the first time
\end{itemize}

\subsection{Cultural factors}

\begin{figure}
\includegraphics[width=70mm, scale=0.55]{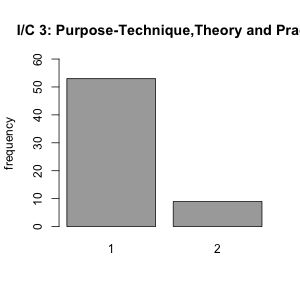}
 \caption{1=Agree,2=Disagree}
\end{figure}
\textbf{Individualist and Collectivist  (I/C) measures(Q33,Q41,Q34)}\\

 In this study there were 3 questions that measured individual/collective culture. Most participants indicated that student would not be embarrassed to come up with new ideas in class (somewhat likely 35\%(n=22))(Question 33; fig 9), having a diploma increases self-respect (strongly agreed 6\%(n=4) and agreed 37\%(n=23))(Question 41; fig 10), and agreeing that the purpose of the lesson was to teach students techniques, practice and theory (85\%(n=53))(Question 34; figure 8).\\
 
\begin{figure}
\includegraphics[width=70mm, scale=0.5]{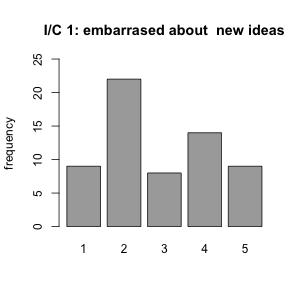}
\caption{1=Very likely, 2=Somewhat likely,3=Neither,4=Somewhat unlikely, 5=Very unlikely}
\end{figure}

\begin{figure}
\includegraphics[width=70mm, scale=0.5]{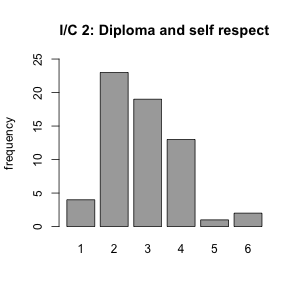}
\caption{1=Agree strongly, 2=Agree,3=neither,4=Disagree, 5=Disagree strongly, 6=not reported}
\end{figure}

\textbf{Power distance (PD) measure (Q30,Q31,Q32,Q40)}\\
There were 4 questions that measured power distance aspect of culture. Most participants indicated that teachers are the most important resource of learning inside classroom(agreed 89\%(n=55))(Question 30; fig 11), teachers determine learning style in a classroom (agreed 81\%(n=50))(Question 31; fig 12), students are hesitant to ask questions because they think teacher is more intelligent (very unlikely 24\%(n=15))(Question 32; fig 13), students respecting teachers outside of classroom (agreed 48\% (n=30))(Question 40; fig 14). \\

\begin{figure}
\includegraphics[width=70mm, scale=0.55]{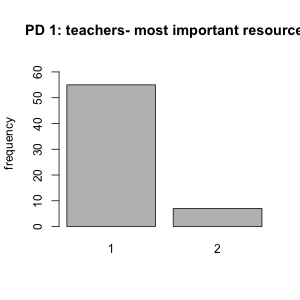}
\caption{1=Agree,2=Disagree}
\end{figure}

\begin{figure}
\includegraphics[width=70mm, scale=0.5]{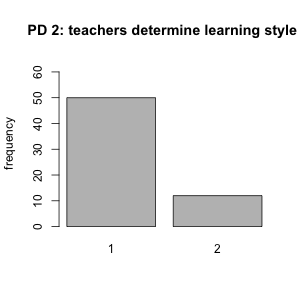}
\caption{1=Agree,2=Disagree}
\end{figure}

\begin{figure}
\includegraphics[width=70mm, scale=0.5]{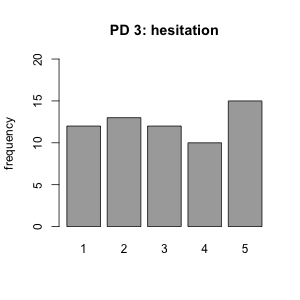}
\caption{1=Very likely, 2=Somewhat likely,3=Neither,4=Somewhat unlikely, 5=Very unlikely}
\end{figure}

\begin{figure}
\includegraphics[width=70mm, scale=0.5]{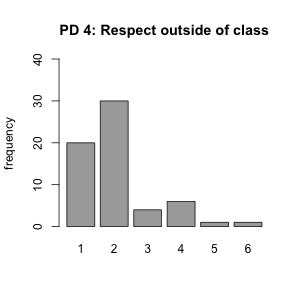}
\caption{1=Agree strongly, 2=Agree,3=neither,4=Disagree, 5=Disagree strongly}
\end{figure}

\textbf{Masculine and Feminine (M/F) measure  (Q35,Q36,Q37,Q42)}\\
There were 4 questions that measured masculine/ feminine culture. Participants indicated that interaction is friendly outside of classroom (very friendly(11\%(n=7)),somewhat friendly(24\%(n=15))(Question 35; fig 15), the presence of physical aggression in classroom (disagreed 47\%(n=29),(strongly disagreed 25\%(n=16)))(Question 36; fig 17 ), students challenging teachers intellectually (strongly agreed 11\%(n=7), agreed 24\%(n=15), did not answer 6\%(n=4)) (Question 37; fig 16), failing of a year (strongly disagreed 35\%(n=22), disagreed 21\%(n=13))(Question 42)\\

\begin{figure}
\includegraphics[width=70mm, scale=0.5]{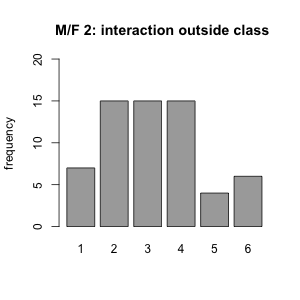}
\caption{1=Very friendly, 2=Somewhat friendly,3=Neither,4=Somewhat unfriendly, 5=Very unfriendly, 6=unanswered}
\end{figure}
 
 \begin{figure}
\includegraphics[width=70mm, scale=0.5]{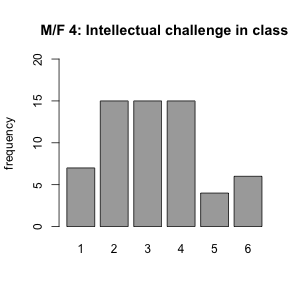}
\caption{1=Agree strongly, 2=Agree,3=neither,4=Disagree, 5=Disagree strongly,6=unanswered}
\end{figure}

 \begin{figure}
\includegraphics[width=70mm, scale=0.5]{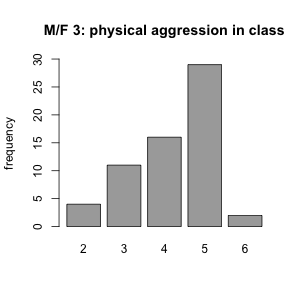}
\caption{2=Agree, 3=neither, 4=Disagree, 5=Disagree strongly, 6=unanswered}
\end{figure}

\textbf{Uncertainty Avoidance (UA) measure (Q38,Q39,Q43)}\\
There were 3 questions that measured UA. Participants indicated that students were comfortable learning in a structured learning environment (strongly agreed 55\%(n=34), agreed 34\% (n=21), no one strongly disagreed)(Question 38; fig 19), teachers\textquotesingle importance of complementing students\textquotesingle accuracy of work (very important 76\%(n=47))(Question 39; fig 20), I do not know the answer(strongly agreed 32\%(n=20), agreed 32\%(n=20)) (Question 43; fig 18) \\

\begin{figure}
\includegraphics[width=70mm, scale=0.5]{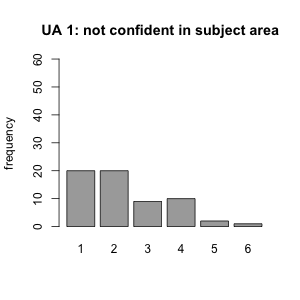}
\caption{1=Agree strongly, 2=Agree,3=neither,4=Disagree, 5=Disagree strongly,9=unanswered}
\end{figure}

\begin{figure}
\includegraphics[width=70mm, scale=0.5]{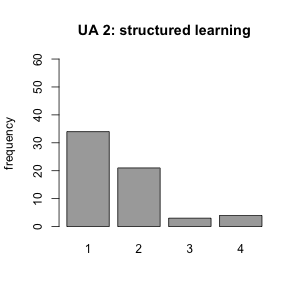}
\caption{1=Strongly agree,2=Agree, 3= Neutral, 4= Disagree}
\end{figure}

\begin{figure}
\includegraphics[width=70mm, scale=0.5]{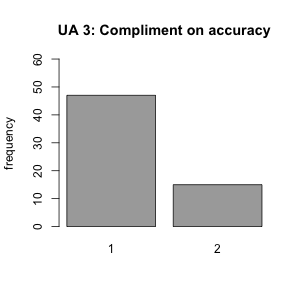}
\caption{1=Agree,2=Disagree}
\end{figure}

\subsection{Grit factors}
This section questioned respondents about their  non-cognitive measure, grit (passion and perseverance) to complete tasks in order to measure their intrinsic motivation.\\

\textbf{Passion(Q44,Q46,Q48)}\\
Participants indicated to the item,\lq I often set a goal but later choose to pursue a different one\rq  (Question 48)attained the maximum mean score 3.9, \lq My interests change from year to year \rq(Question 46) had a mean score 3.6 and \lq Sometimes new projects and ideas distract me from previous ones\rq (Question 44) had a mean score  3.6 (fig 21).\\   

\begin{figure}
\includegraphics[width=70mm, scale=0.5]{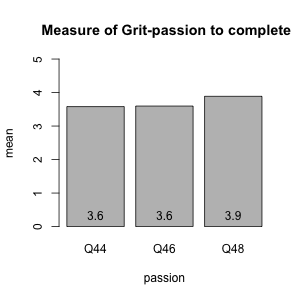}
\centering
\caption{Intrinsic motivation- passion. Refer to questionnaire in appendix}
\end{figure}

\textbf{Perseverance(Q45,Q47,Q49)}\\

Participants indicated to the question of \lq I work hard \rq (Question 45) attained the maximum mean score 4.3,\lq  I finish whatever I begin \rq (Question 47) had a mean score of 4.0 and \lq setbacks don\textquotesingle t discourage me. I don\textquotesingle t give up easily\rq had a mean score of 4.0 (Question 49)(fig 22).\\  

\begin{figure}
\includegraphics[width=70mm, scale=0.5]{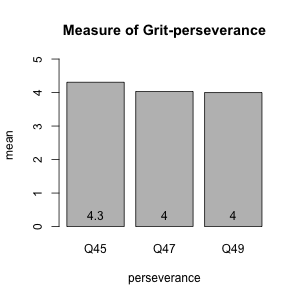}
\centering
\caption{Intrinsic motivation- perseverance. Refer to questionnaire in appendix}
\end{figure}

\begin{figure}
\includegraphics[width=70mm, scale=0.5]{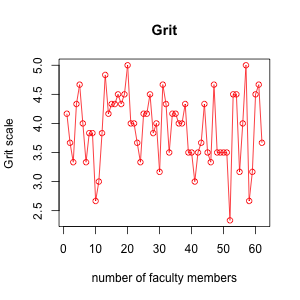}
\caption{Respondents' Grit measure }
\end{figure}

\subsection{Open questions}
In a series of open-ended questions, HE staff were asked about their personal views about learning technology and achievement. Below are five randonly selected responses from participants. \\
\textbf{What is your personal view in using technology in learning?}\\

\textbf{Respondent 1:}\\
We have to use technology, there is no option as such. If learning is to happen in the concurrent world whatever technology is there has to be used. It is very much related to communication. Without using technology, no learning is possible. We have dumped old overhead projectors in the corner of the office, today it can be used but people will not appreciate it. Now there are new technologies and we have to go by that.\\

\textbf{Respondent 2:}\\
Technology in learning is inevitable in a way. We go by and everything we do will be based on technology and there will be virtual classrooms and people will start learning from home. Teachers should be ready to imbibe such technologies and be prepared to teach through these e-learning platforms just by sitting at home or office, virtually 0\% students in front of them. That will come and is not too far. That will happen very soon.  

\textbf{Respondent 3:}\\
Very positive view. Technology has to come up. It is really important, nowadays we have online teaching but I don\textquotesingle t really do online. When we compare between both, I very much prefer online. That personal touch will not be there in online learning  but in face-to-face we will  be able to assess the body language of the student. We will be able to divert the attention of the student and we will be able to better entertain the student. Where as in online teaching it will be just a matter of fact teaching. I prefer direct classes but at the same time, for delivering the assignments technology should be there. \\

\textbf{Respondent 4:}\\
I very much support technology in learning and there are reasons why I believe. To some areas yes, obviously automation can and is happening a lot but I think in management education the basics I don\textquotesingle t think without a teacher it will be easy for a student to get it. Because most of the books are also  written in such a way that not everyone will understand. So I am not talking about  the top gear or the cream students; my working experience is with students who are not taught much so I think at their level a teacher is indispensable.\\

\textbf{Respondent 5:}\\
I think technology is vital. Technology is a good thing to have while you are doing teaching and learning. I don\textquotesingle t think teacher can be replaced. I feel that teacher is a point to provide feedback. You can learn whatever you want from a video but you cannot know whether you have learned it. I know you  can take exams and things like that when you are learning but you want someone to give you feedback.\\

\textbf{How would you describe achievement?}\\

\textbf{Respondent 1:}\\
If you can achieve 80\% of your goal, then that is achievement. Reaching 80\% of target you have set may be a good achievement. In our set up we have teaching, research and service. So in the research area may be everything is weighted with projects and  publications with  national and international societies, targeting various positions of office bearers. Position in the society as an office bearer matters.\\

\textbf{Respondent 2:}\\
According to me , when I set a target if I am able to achieve that then that is achievement for me. Everyday is quiet challenging. No two days are same. As I enter the class, everyday is a challenge for me how to convince the students. How to put my point across and if I have done it then the day is mine. You should be successful not just for exams. I want to prepare them not just for exam but for their life.\\

\textbf{Respondent 3:}\\
Achievement is something that provides you satisfaction. Actually, it comes when you balance work with your personal priorities. Should be able to balance both personal interest and professional interest. As a teacher I would come up with good publications, right now I have x number of publications and I want to publish in more internationally best journals. That is one thing I should focus now. The rest of the things I am happy.\\

\textbf{Respondent 4:}\\
It depends on the number of students that we have taken, taken them to a state were they could achieve. Achievement of every student is the achievement of the teacher. The teacher does not have any special achievement other than this.\\

\textbf{Respondent 5:}\\
I can say that I have achieved something if my colleagues, students or younger colleagues  have imbibed my interests when I was dealing with them. They work in that field and it comes to a level that when I go nothing happens to that concept. Influence in the sense, create interest in a topic, that is the final duty of a teacher. If you have created it then students will take it over. University is to bring about knowledge, this is to be continued by people. My students write 100 papers on the topic that I have been teaching them  starting with me, then that is my achievement.\\

\section{Discussion}
Most of the participants of the survey were from private-unaided institutions. The cohort studied was experienced in teaching(mean teaching experience 11 years) and they  had tried different teaching methods including some degree of electronic learning. Indeed all respondents used electronic resources for preparing for lessons however only half of the cohort used the internet during the lesson.  The types of resources used from the internet were most commonly  PDFs and e-books. Slightly more than half of institutions had invested in developing their learning platforms. \\It seems that the challenges faced by staff are many.\\

The results of this study highlights several challenges faced by staff in the HE sector across India to implementing education technologies. These include the  lack of dedicated time set aside for maintaining e-learning systems up-to-date, cultural factors and lack of passion. When evaluating the cultural measure of \textit{Individualistic/Collective}(I/C), this cohort exhibited a mix of responses supporting both individual and collective natures; however the results suggest that the lecturers preferred a collective culture of looking after the group\textquotesingle s interest  rather than the individual \textquotesingle s interests. This collective culture is an element to be taken into consideration when implementing  technology frameworks.The \textit{Power Distance} (PD)measure  represents the inequality in society and evaluating the emotional distance that separates subordinates from their bosses, this study indicated markers that were suggestive of  cultures with greater Power Distances. A hierarchal model between teacher and student is still strongly embedded in Indian culture and educational technology should incorporate this to be successfully implemented. The \textit{Masculine/Feminine}(M/F) measure which reflects the emotional gender roles in the mental programming in societies, and it was found that the responses were mainly masculine.  
Lastly, evaluating  the culture measure of \textit{Uncertainty Avoidance}(UA),  which is the extent to which the members of a culture feel threatened by ambiguous or unknown situations, the result suggest that HE staff prefer a culture where there is less  ambiguity and uncertainty. Any educational technology to be implemented in such cultures should be aware of this and avoid  uncertainties where possible. The lecturers in this cohort were very gritted professionals who had greater perseverance than passion, which suggests that with the culturally appropriate frame work the cohort are intrinsically motivated to complete challenging tasks.\\

\section{Conclusion}
This study sought to explore a culturally adaptive learning environment which is appropriate to the requirements of educational technology, and the result that reflected positive use of technology among lecturers in terms of lesson preparation and delivery and also gave different cultural and motivational indications about the role of an educational technology that is culturally adaptive. \\
Our study has shown that cultural challenges are significant and educational environments are affected by the cultural influences in the context of  historical background of a country.  Users in a culture which are different from western culture require a learning technology that integrates with their culture. A culturally appropriate learning platform will contribute to the creation of a workforce relevant to that particular society and would be an effective way to converge the motives of government, businesses and local employment.
\section{acknowledgement}
The author acknowledges Dr. James Ohene-Djan, Goldsmiths, University of London for input received during the design of the study. 

\bibliography{paper3}
\bibliographystyle{apalike}

\appendix

\section{Questionnaire}
Q1. How many days in a week do you teach?\\
Q2. What is the teaching mode ? \\
Q3. Do you access electronic resources while preparing for your lesson? (select one) Yes/No\\
Q4. If yes, Which of the following learning activities do you do while preparing for the lesson? (Please select all applicable)\\Q5(i)Worked on your own through materials 			
online (pdfs, videos, Wikipedia)\\
Q6(ii)Worked on your own through materials 			
on CD-ROM\\
Q7(iii)Read an e-book\\
Q8 (iv)Worked on your own through a package
of materials online which lead to 
a qualification in the last 10 years\\
Q9(v)Worked on institutional 
learning platform such as
Moodle/in-house build\\
Q10. Do you use internet while you teach? (highlight one)(Yes/No)\\
If yes, On average how often do you use the following in a week? (Please highlight all applicable)\\
Q11. I.	Searching the internet\\
5 days 	4 days	3 days	2 days	1 day	none\\
Q12. II.	Following specific learning courses online	\\
5 days 	4 days	3 days	2 days	1 day	none\\
Q13. III.	Downloading information from the web to read\\
5 days 	4 days	3 days	2 days	1 day	none\\
Q14. IV.	Surfing an intranet between Institutions\\
5 days 	4 days	3 days	2 days	1 day	none\\
Q15. V.	Lessons to student(s) via other applications such as Skype\\
5 days 	4 days	3 days	2 days	1 day	none\\
Which of the following problems do you experience when using technology for teaching? (Please highlight all applicable)\\
Q16.  Appropriate hardware tools\\
Q17. Appropriate software tools\\ 
Q18. Bandwidth (data transfer rate)\\
Q19. Competency with English language\\ 
Q20. Time to use e-learning systems properly\\
Cost \\
Q21. Curriculum\\
Q22.Institute?s policy implementation\\
Q23.Internet connectivity \\
Q24.Power supply\\
Q25.Students\textquotesingle technological confidence\\ 
Q26.Technological expertise by staff\\
Q27.Learning curve\\
Q28.Keeping e-learning system up-to-date\\
Q29. Any others\\
Q30. Do you agree or disagree that teachers are the most important resource of learning inside the classroom(Agree/Disagree)\\
Q31. Do you agree or disagree that teachers should determine the learning style and pace for students during a lesson	Agree/Disagree)\\
Q32. How likely do you think your students are hesitant to ask questions because they think you are more intelligent than them?(very likely/somewhat likely/neither likely or unlikely/somewhat unlikely/very unlikely)\\
Q33. How likely or unlikely do you think your students are embarrassed to come up with new ideas in class?(very likely/somewhat likely/neither likely or unlikely/somewhat unlikely/very unlikely)\\
Q34. Do you agree or disagree that the purpose of your lesson is to teach your students how to do techniques, practice and theory ?Agree/Disagree)\\
Q35. How friendly is your interaction with students outside of classroom? (very friendly/ somewhat friendly/ neutral/somewhat unfriendly/ very unfriendly)\\
Q36. To what extent do you agree or disagree that students are physically aggressive in class?(strongly agree/agree/ neutral/disagree/strongly disagree )\\
Q37. To what extent do you agree that students challenge the teacher intellectually? (strongly agree/agree/neutral/disagree/strongly disagree)\\
Q38. How comfortable or uncomfortable are students while learning in a structured environment(very comfortable/somewhat comfortable/neither comfortable or uncomfortable/somewhat uncomfortable/very uncomfortable)\\
Q39. How important do you feel students\textquotesingle should be complimented for accuracy of their work (very important/ somewhat important/ neither important or unimportant/ somewhat unimportant/ very unimportant)\\
Below are several statements regarding societal attitudes. Please read each one and indicate the extent to which you agree or disagree with each statement by selecting the appropriate box for each statement.\\
Q40.Students respect teachers even outside class.(disagree strongly/disagree/neither agree or disagree/ agree/agree strongly)\\
Q41. Having a diploma increases self- respect. (disagree strongly/disagree/neither agree or disagree/ agree/agree strongly)\\
Q42. Failing a year is a minor incident. (disagree strongly/disagree/neither agree or disagree/ agree/agree strongly)\\
Q43. It is alright for teachers to say ?I don?t know?. (disagree strongly/disagree/neither agree or disagree/ agree/agree strongly)\\
Below are several statements regarding you. Please read each one and indicate the extent to which you agree with each statement by marking the appropriate box for each.\\
Q44. Sometimes new projects and ideas distract me from previous ones. (not at all like me/ not much like me/some-what like me/mostly like me/very much like me)\\
Q.45. I work hard. (not at all like me/ not much like me/some-what like me/mostly like me/very much like me)\\
Q46. My interests change from year to year. (not at all like me/ not much like me/some-what like me/mostly like me/very much like me)\\
Q47. I finish whatever I begin. (not at all like me/ not much like me/some-what like me/mostly like me/very much like me)\\
Q48. I often set a goal but later choose to pursue a different one.(not at all like me/ not much like me/some-what like me/mostly like me/very much like me)\\
Q49. Setbacks don\textquotesingle t discourage me. I don\textquotesingle t give up easily. (not at all like me/ not much like me/some-what like me/mostly like me/very much like me)\\
Q50. What is your personal view about using technology in learning ?\\
Q51. How would you describe achievement? \\
\end{document}